\documentclass{article}

\usepackage{arxiv}

\usepackage[utf8]{inputenc} 
\usepackage[T1]{fontenc}    
\usepackage{hyperref}       
\usepackage{url}            
\usepackage{booktabs}       
\usepackage{amsfonts}       
\usepackage{nicefrac}       
\usepackage{microtype}      
\usepackage{cleveref}       
\usepackage{lipsum}         
\usepackage{graphicx}
\usepackage{natbib}
\usepackage{doi}

\usepackage{tabularx}

\usepackage{multirow}
\usepackage{diagbox}
\usepackage{csquotes}

\usepackage{textcomp}

\title{Computational Reproducibility in Computational Social Science}

\author{
	\href{https://orcid.org/0000-0003-2952-4812}{\includegraphics[scale=0.06]{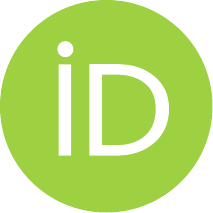}\hspace{1mm}David
	Schoch} \footnote{Correspondence concerning this article should be addressed
	to David Schoch (david.schoch@gesis.org)}\\
	GESIS - Leibniz Institute for the Social Sciences \\
	\texttt{david.schoch@gesis.org} \\
 	\And
    \href{https://orcid.org/0000-0002-6232-7530}{\includegraphics[scale=0.06]{orcid.pdf}\hspace{1mm}Chung-hong
    Chan} \\
	GESIS - Leibniz Institute for the Social Sciences \\
	\texttt{chung-hong.chan@gesis.org} \\
  	\And
    \href{https://orcid.org/0000-0002-0640-8221}{\includegraphics[scale=0.06]{orcid.pdf}\hspace{1mm}Claudia
    Wagner} \\
    GESIS - Leibniz Institute for the Social Sciences \\
	\texttt{claudia.wagner@gesis.org} \\
    \And Arnim Bleier \\
    GESIS - Leibniz Institute for the Social Sciences \\
	\texttt{arnim.bleier@gesis.org} \\
 }

\renewcommand{\shorttitle}{Computational Reproducibility in CSS}

\begin{document}
\maketitle

\begin{abstract}
Replication crises have shaken the scientific landscape during the last decade.
As potential solutions, open science practices were heavily discussed and have
been implemented with varying success in different disciplines. We argue that
computational-x disciplines such as computational social science, are also
susceptible for the symptoms of the crises, but in terms of reproducibility. We
expand the binary definition of reproducibility into a tier system which allows
increasing levels of reproducibility based on external verfiability to
counteract the practice of open-washing. We provide solutions for barriers in
Computational Social Science that hinder researchers from obtaining the highest
level of reproducibility, including the use of alternate data sources and
considering reproducibility proactively.   

\end{abstract}

\keywords{Computational Reproducibility \and Computational Social Science}
\section{Defining reproducibility}

The 2010s have been referred to as the decade of the replicability crisis in
psychology~\citep[]{nosek:2022:RRR,OSC2015:E} as well as other social sciences
\citep[e.g.][]{wuttke:2018:WTM}. The replicability crisis denotes the phenomenon
that many scientific studies cannot be replicated or confirmed by independent
researchers~\citep{shrout2018PSK}. A reaction to this crisis was a growing
movement in the scientific community to improve research practices and increase
transparency in reporting scientific results \citep{mnbbcpswwi-mrs-17} such as
pre-registering studies to reduce bias, making data and code openly available,
and conducting more replication studies. Many of the proposed \emph{enablers of
replicability} can be summarized under the term ``Open
Science''~\citep{vm-osnslrid-18}. This call was perhaps complicated by the fact
that at the same time computational methods have begun to be applied in studying
social phenomena \citep{lazer2009} which led to the rise of computational social
science (CSS).  CSS research includes research on almost every topic related to
human behavior. It is thus not a discipline centered around a single branch of
knowledge, like knowledge about the brain in neuroscience. Rather, it is
characterized by data, and computational and statistical methods used to provide
evidence-driven answers to questions that emerge in other social scientific
disciplines. In this article, we argue that it is this data and method focus
that makes CSS susceptible for the symptoms of the described crises. Not in
terms of replicability, but reproducibility.

Since it is difficult to discuss the reproducibility of CSS research using its
current binary definition, we thoroughly define the term computational
reproducibility in our article. Building upon this conceptual clarity, we define
common barriers to achieving computational reproducibility and recommendations
on how to resolve them.

\subsection{A Tier System of Computational Reproducibility}

There has been a great deal of confusion about the meanings of ``replicability''
and ``reproducibility''. The two terms are sometimes even used interchangeably
and different fields have different understandings of both
\citep{barba2018terminologies}. We initialize our discussion with the
definitions provided by the American Statistical Association (ASA) definition of
replication: ``the act of repeating an entire study, independently of the
original investigator without the use of original data (but generally using the
same methods).'' Reproducibility, is defined by the ASA as ``you can take the
original data and the computer code used to analyze the data and reproduce the
numerical findings from the study.'' The distinction between replicability and
reproducibility, together with robustness and generalizability, is commonly
presented as a two-by-two grid like in Table \ref{tab:2by2}.

\begin{table}[ht]
    \centering
    \caption{Two-by-two grid defining reproducibility and replicability.}
    \label{tab:2by2}
    \begin{tabular}{ll|ll}
        &&\multicolumn{2}{c}{\textbf{Data}} \\
        && Same & Different \\
        \hline
        \multirow{2}{*}{\textbf{Analysis}} &Same & Reproducible & Replicable \\
        &Different & Robust & Generalisable \\
    \end{tabular}
\end{table}
A drawback of this two-by-two presentation is that it does not present the
criteria of a successful event. A definition including such criteria is given
by~\citet{The_Turing_Way:2022}: ``A result is reproducible when the
\textit{same} analysis steps performed on the \textit{same} dataset consistently
produces the \textit{same} answer.'' Note that this definition is compatible
with the ASA definition. 

We argue that both definitions are insufficient to describe computational
reproducibility unambiguously since they both neglect two dimensions that are
important for its assessment: (i) ``who is the agent that is able to conduct the
reproducibility check?''  and (ii) ``what is the computational environment in
which the reproducibility check can be conducted?'' Existing definitions for
reproducibility lack clarity about those dimensions. The Turing Way definition
does not specify the agent, whereas the definition by the ASA uses an ambiguous
``you''. Without this ambiguity being clarified, ``you'' may refer only to
the original investigators (\emph{first party }), all other researchers
(\emph{third party}) and/or trusted institutions such as journals that may have
verified the reproducibility of the results that they publish (\emph{trusted
third party}).

Another drawback of the Turing Way and the ASA definition of reproducibility is
that they do not consider the computational environment at all, despite its
importance for computational reproducibility. We argue that the accessibility of
suitable computational environments should be treated as equally important as
the accessibility of data and methods. Reproducible materials \footnote{Center
for Open Science define (reproducible) materials as: Components of the research
methodology needed to reproduce the reported procedure and analysis.} should
expand and include the information on the computational environment. Therefore,
the computational environment becomes an integral part of the definition for
computational reproducibility. 

In our article we differentiate between the \emph{private materials} that can
only be used by the original authors, \emph{restrictive reproducible materials
for trusted third parties} and \emph{nonrestrictive reproducible materials}.
When being shared, nonrestrictive reproducible materials should contain
information to allow everyone to consistently generate a computational
environment in which the original code and data can be loaded and executed.
Specifically, this means that the material should not access anything that does
not already belong to the materials such as external APIs or data.  Restrictive
reproducible materials allow only selected third parties e.g. those with access
to sensitive data or who have special computational equipment or proprietary
software, to generate an environment in which data and code can be executed and
produce consistently the same results proposed by the original investigator(s).
Finally, private materials are assumed to be not shared at all and are
consequently not reproducible.

We define the basic level of computational reproducibility as ``A result is
reproducible when the \textit{same} analysis steps performed on the
\textit{same} dataset consistently produce the \textit{same} answer. This
reproducibility of the result can be checked by the original investigators and
other researchers with the nonrestrictive reproducible materials.'' To ease the
discussion in the rest of this paper, we call this first-order computational
reproducibility (1\textdegree CR) which allows us to define a tier system of
computational reproducibility. This basic level of computational reproducibility
is not verified externally. However, the original authors have given all the
materials, including data, code, and information about the computational
environment, to third-party agents for verification.

A result can be said to be \textbf{externally verified} computational
reproducible, or having third-order computational reproducibility (3\textdegree
CR) when the \textit{same} analysis steps performed on the \textit{same} dataset
by \textit{external researchers} consistently produce the \textit{same} answer.
This involves the execution of the shared computer code with the shared data by
external researchers. 

For practical reasons, we also define the state of second-order computational
reproducibility (2\textdegree CR) where only trusted third-party agents can
confirm the computational reproducibility. This limited state is sometimes
useful when a finding is based on the analysis of highly sensitive data or with
highly specialized equipment. This state is constrained by the restrictive
access to materials such as data or computational resources. The complete tier
system of computational reproducibility is summarized in Table
\ref{tab:execution}.

\begin{table}
    \centering
    \caption{\textbf{Execution Matrix.} This table shows the tiers of computational reproducibility based on the agent that conducts the reproducibility test and the reproducible materials that the agent uses for the test. 
    }
    \label{tab:execution}
    \begin{tabular}{|l|p{2,7cm} |p{2,7cm} |p{2,7cm} |}\hline
     \toprule
     \diagbox[width=14em]{\textbf{Agent}}{\textbf{Materials}}&
      \textbf{Private}&
      \textbf{Restrictive}& 
      \textbf{Nonrestrictive}\\
  \midrule
      
      \textbf{Author(s) only} & 
      --- &
      --- &
      1\textdegree CR\\ \hline
  
      \textbf{Trusted \mbox{third-parties}} & 
      --- &
      2\textdegree CR & 
      3\textdegree CR \\ \hline
  
      \textbf{Everyone} & 
     ---  &
     ---  & 
      3\textdegree CR\\
  
      \bottomrule
      \end{tabular}
  
  \end{table}
  

The purpose of this tier system is to improve the scientific discourse about
computational reproducibility and counteract the practice of open-washing (i.e.,
making research look reproducible only for marketing purposes).

\section{Barriers Imposed by External Dependencies and Opacity} %

External dependencies refer to the parts in the research pipeline that depend on
external entities (e.g. APIs or external libraries). External dependencies are a
barrier to computational reproducibility when researchers have no control over
the external entities and/or when external entities are not transparent.

Prominent examples of external dependencies that fall into both classes (no
control and not transparent) are APIs for data access and data analysis (such as
the Twitter API, ChatGPT or  Perspective API). Materials with dependencies on
external entities that are not under the control of the researcher and are not
transparent cannot be considered computational reproducible since the behavior
and functionality of the external entity can change at any point in time, as
well as its accessibility.

 The most recent example of a sudden change in accessibility is the deprecation
of free academic API from Twitter in 2023. This renders much existing research
irreproducible: Shared Twitter datasets in the permissible format allowed by X
Corp. (formerly Twitter Inc.) are Tweet IDs. These shared Tweet IDs are now
meaningless because retrieving a tweet's complete information using its ID is
not feasible anymore \citep{assenmacher:2023:ERE}.

The ups and downs of API access has left CSS flip-flopping between a ``Golden
Age of Data''~\citep[]{grady2019golden} and a ``Post-API
age''~\citep{freelon:2018:CRP}. However, even without restrictions in place,
using such data can pose barriers for reproducibility efforts. APIs have always
been proprietary black boxes, which were never really intended for academic
use~\citep{t-whadgcradrpa-21}, and we rarely know what type or quality of data
we are analyzing~\citep{mplc-sgecdtsatf-13}. If data explicitly needs to be
gathered again to reproduce results, then we cannot guarantee that all relevant
data is still available.  Moreover, even when the data is still available, the
terms of use of data gathered from the API may have changed and make certain
research impossible. For instance, recently Reddit decided to not allow training
machine learning models with their data anymore~\citep{davidson2023}. These
issues render original results potentially irreproducible without access to the
raw data.

Besides data access, external dependencies on big tech companies also threaten
other aspects of the research pipeline. Many heavily used services by
researchers to foster computational reproducibility are owned by big tech
companies, for instance, Colaboratory (Google) and GitHub (Microsoft). As with
API access, it is imprudent to assume that these companies will maintain the
provision of these services without restrictions. Google and Microsoft also have
a history of discontinuing popular research services such as Google Fusion Table
and Microsoft Academic. This potentially renders all research studies using
these products obsolete along with them. Therefore, one should consider such
services will expire when the generosity of these big tech companies runs out.

A last barrier relating to external dependencies concerns the use of proprietary
commercial software such as Stata and Tableau, which impose a preventable
economical barrier to reproducibility: external researchers are obliged to
purchase licenses just for the purpose of verifying the reproducibility of a
result generated from these proprietary software tools. This issue also manifests in
another way by applying proprietary services in the form of RESTful APIs, e.g.
ChatGPT, Perspective API or Botometer \citep{yang2022botometer}. Scholars have
argued that these ``blackbox'' services are intrinsically irreproducible because
the algorithms running at the service provider side are constantly changing
\citep[]{rauchfleisch:2020:F,chen2023chatgpt}. Similarly, the aforementioned
economical barrier applies, if these services (e.g. ChatGPT) carry a cost.

\subsection{Solutions: Alternative Data Sources and Open Source Software}

External dependencies and the lack of transparency are tremendous threats to
computational reproducibility and should be avoided at all cost. 

On the data front, a possible alternative is to shift from relying on the
generosity of big tech to the generosity of users, via digital data
donations~\citep[]{ohme:2023:DTD}. Another promising direction is setting up
legal frameworks that empower industry-research collaborations and ensure free
access to industry data by researchers (such as the Digital Service Act).

If APIs from these big-tech companies must be used, we recommend caching
complete output from these APIs as data files. Solutions such as vcr \citep{vcr}
can be used. We can then consider the API outputs as some static sensitive data.

Finally, researchers should consider the option to archive their data (the
cached static API outputs and beyond) in a secure data archive\footnote{e.g.
\url{https://www.gesis.org/angebot/daten-aufbereiten-und-analysieren/analyse-sensibler-daten/secure-data-center-sdc}}
that will ensure the long-term accessibility of data for other researchers (if
necessary under certain contracts). 
Many papers might have the claim that data and computer code are available upon
request. However, several previous studies have shown that researchers do not
necessarily respond to these requests \citep{vaadbfgmrr-arddraa-14}, or they
themselves do not have access to their own data and code as time goes by
\citep[e.g.][]{tedersoo:2021:D}. \cite{vaadbfgmrr-arddraa-14} report that the
availability rapidly decreases with age of the paper. However, even if journals
have stricter mandates on data and code sharing, they seemed to be ignored in
the past \citep{vabfgkmmrrvy-mdagiard-13}. In contrast, if papers come with a
data (and code) availability statement, 80\% of data and code remain available
even over a longer period of time \citep{f-ladaapo-22}.

On the analysis front, we advise against the release of academic research
software in the form of RESTful API, as in the example set by the widely used
Botometer~\citep[][now defunct due to the closure of Twitter
API]{yang2022botometer}. Maintenance of these APIs requires enormous amount of
computational resources and the reproducibility of downstream research using
these APIs depends on the sustained allocation of these resources. Researchers
should consider not using commercial API services such as ChatGPT in their CSS
research. When free and open source alternatives are available, those
alternatives should be considered first. For example, prompt-based LLMs such as
ChatGPT can be replaced with Alpaca~\citep[]{alpaca} \footnote{Having said so,
open source LLMs are difficult to define, depending on factors such as whether
the training data are open and whether the license is restrictive. For a guide
on how to choose an appropriate open source model, see
\citep[]{liesenfeld:2023:OC}.}. These locally deployed software might not have
the state-of-the-art performance. But that performance impact ---usually not
important enough for the analysis and can be statistically
adjusted~\citep[]{teblunthuis2023misclassification} --- is a reasonable price to
pay to improve reproducibility. One often ignored aspect of using these
commercial API services is that private or sensitive data often get transferred
to a commercial entity (e.g. OpenAPI), to whom the participants of the study
probably did not give consent for researchers to make the data transfer. 

In the same vein, researchers should use open source software as much as
possible. For CSS research, this is relatively easy due to the \textit{de facto}
duopoly of R and Python as the most used programming languages
\citep[]{metzler2016doing}. 

\section{Barriers Imposed by Computational Aspects}

Computational barriers for reproducibility can be divided into human and
technical components. The human induced barriers include the still high
unwillingness to actually share~\citep{ch-srcscrpainp-22}, and knowledge gaps in
programming practices, where the former is partially a consequence of the latter
since researchers are afraid that their ``bad'' programming skills are exposed
when sharing code~\citep{ch-srcscrpainp-22}. Other often cited reasons for not
sharing code are time constraints for making the code camera-ready. 

In order to achieve correct and reproducible code, programming knowledge is
needed and knowledge gaps among CSS researchers clearly exist, already due to
the high interdisciplinary of the field~\citep{s-csshais-18}. Beyond
disciplinary boundaries, however, the gap also stems from the fact that the vast
majority of researchers, even those who develop research software, are primarily
self-taught and not equipped with any formal training in software
development~\citep{hannay:2009:H}. Researcher can probably write executable code
but have varying skills in standard software development practices such as using
unit tests and continuous integration~\citep{wabhdghhmpwww-bpsc-14} to show
correctness. Code shared by researchers is usually written to ``do the job
once'' and not with longevity or computational reproducibility in mind.

The diversity of computational environments overshadows the human component.
Previous studies showed that most researchers run their analyses on their
desktop or laptop computers rather than standardized computing environments
\citep[]{hannay:2009:H}. Worse, the details of the computational environment
used in the analysis is usually underdescribed. There are many variations in the
computational environments that can prevent the code from running. As a simple
model and considering only the software layer, \cite[]{chan:2023} list out four
components that can vary from one computational environment to another: (A)
operating system, (B) system components, (C) the exact version of the
programming language, and (D) what and which version of the software libraries.

For instance, many code snippets published in the software paper of the popular
text analysis package quanteda ~\citep[]{benoit:2018} are not executable with
recent quanteda releases due to a major rewrite of the software. One must use a
release prior to April 2021 to run those code snippets. Hence, in this case
issues with component D (exact version of a particular software library) prevent
the code from running.

\subsection{Solutions: Education and Proactive Reproducibility}

As the definition of computational reproducibility calls for the same code and
data \textbf{consistently} produce the same result, one can attain this by
either (1) making sure the code and data produce the same result in all
computational environments, or (2) making sure that there is a way to
consistently generate a compatible computational environment for the code and
data to run on. The former is impossible, while the latter is possible: provide
a reproducible computational environment where at least the four components from
the last section are clearly documented and can be automatically regenerated.

However, not all CSS researchers have the same skill set to achieve such
computational reproducibility. Luckily, there has been existing literature on
how to write reproducible and testable code \citep[e.g.][]{trisovic:2022}. Best
practices for writing reproducible code should be promoted. Inclusive
educational initiatives such as Software Carpentry~\citep{wilson:2006} and The
Turing Way~\citep{The_Turing_Way:2022} should be supported. Also, CSS courses
should focus more on software engineering fundamentals such as software testing.

If the computational reproducibility is not an afterthought (retroactive) but a
built-in feature from the beginning (proactive), the computer code is readily
shareable without cleanup. This proactive approach also allows great room for
automating the usual chores.

Another way to ensure proactive computational reproducibility is to organize
computer code and data of a research project as a reproducible research
compendium. A research compendium encourages researchers to organize computer
code and data separately into a sensible structure. Computer code is documented
and Literate Programming~\citep{knuth:1984} techniques can also be used to make
sure the reporting in the manuscript is perfectly aligned with the data
analysis. Tools have been developed to fully automated the process of authoring
a reproducible research compendium \citep[e.g.][]{schulte:2012:MLC}.

\begin{figure}[h]
    \centering
    \caption{A declarative description of a computational environment (Dockerfile) generated by rang: It pins down the operating system (Debian 5.0 "Lenny"), the system component (libxml2-dev), the R version (2.15.0), and all software libraries (slam, SparseM, Rcpp, tm, maxent) with their versions}
    \includegraphics[scale=0.2]{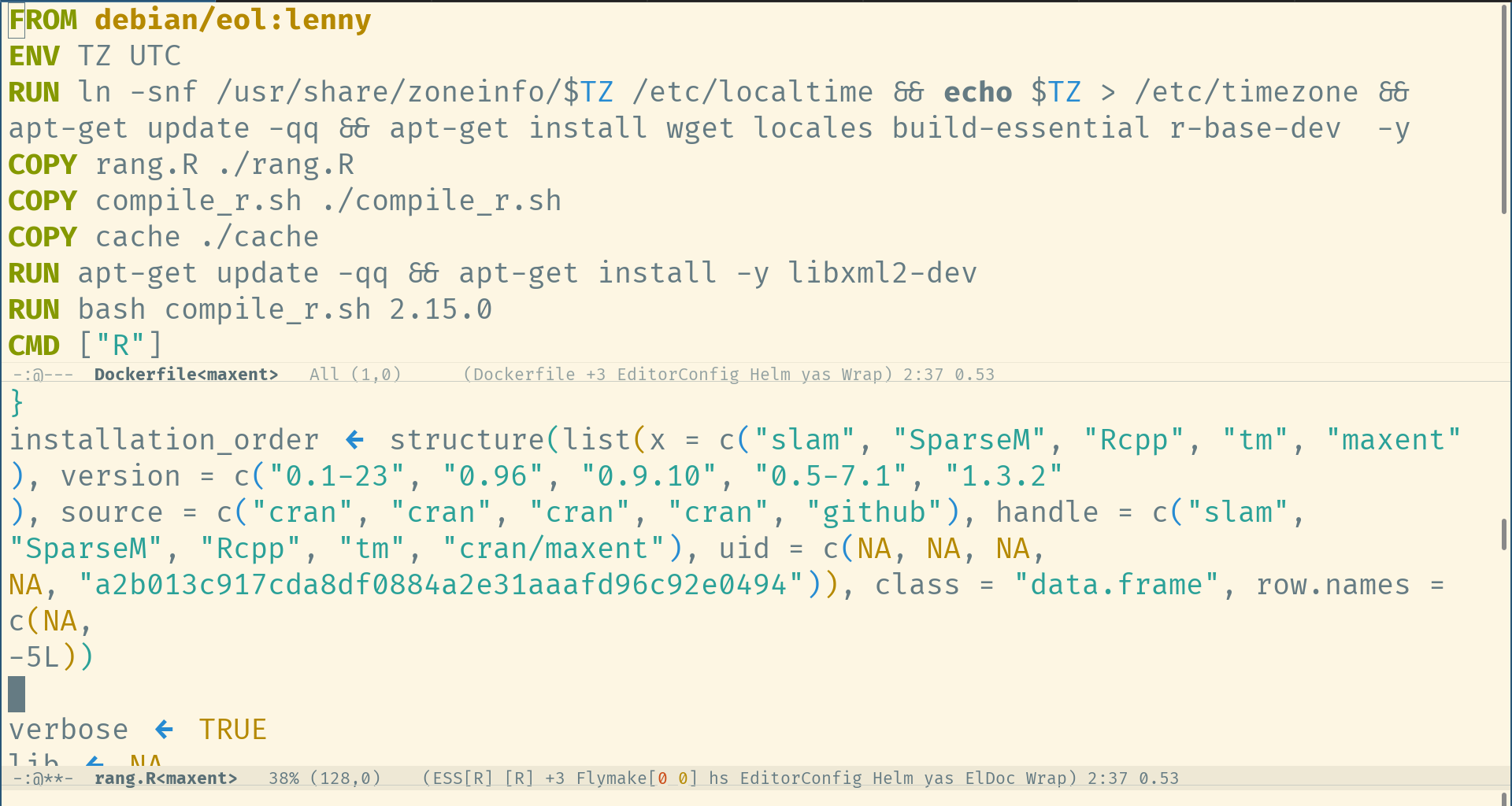}
    \label{fig:rang}
\end{figure}

Virtualization systems, such as Docker and Apptainer, have been recommended to
increase research reproducibility in other fields \citep{wblo-dtrnusp-20}. The
declarative description of a computational environment, such as Dockerfile, is a
plain text file and can be shared together inside a research compendium. The
procedure for writing a Dockerfile can be automated with tools such as
rang~\citep{chan:2023} and repo2docker~\citep{ragan:2018}. Figure \ref{fig:rang}
shows an example of an automatically generated declarative description that pins
down all four software components.

We strongly recommend using Linux-based systems for CSS research, owing to their
compatibility with containerization, and open licensing, all of which enhance
reproducibility. Standard Linux distributions such as Ubuntu are good enough for
reproducibility because the software repositories of these standard Linux
distributions have good archives of their old software packages.

\section{Conclusion \& Recommendations}

In this paper, we introduced a definition for computational reproducibility that
incorporates the agent and computational environment, which allowed us to define
a tier system that covers various levels of verifiable computational
reproducibility based on \emph{who} and \emph{where} results are reproduced. We
argued that most CSS research cannot be considered reproducible given that the
sharing of code and data is still not mandated by many journals and researchers
rarely share their material voluntarily. But even if researchers are willing to
do so, there are two barriers to break through in order to make a piece of
research truly reproducible. For the two barriers, we offered alternative
approaches that facilitate the process of making research verifiably
reproducible.

\begin{figure}[h]
    \centering
    \caption{Recommended practices to achieve the maximum degree of computational reproducibility in different scenarios}
    \includegraphics[scale=0.6]{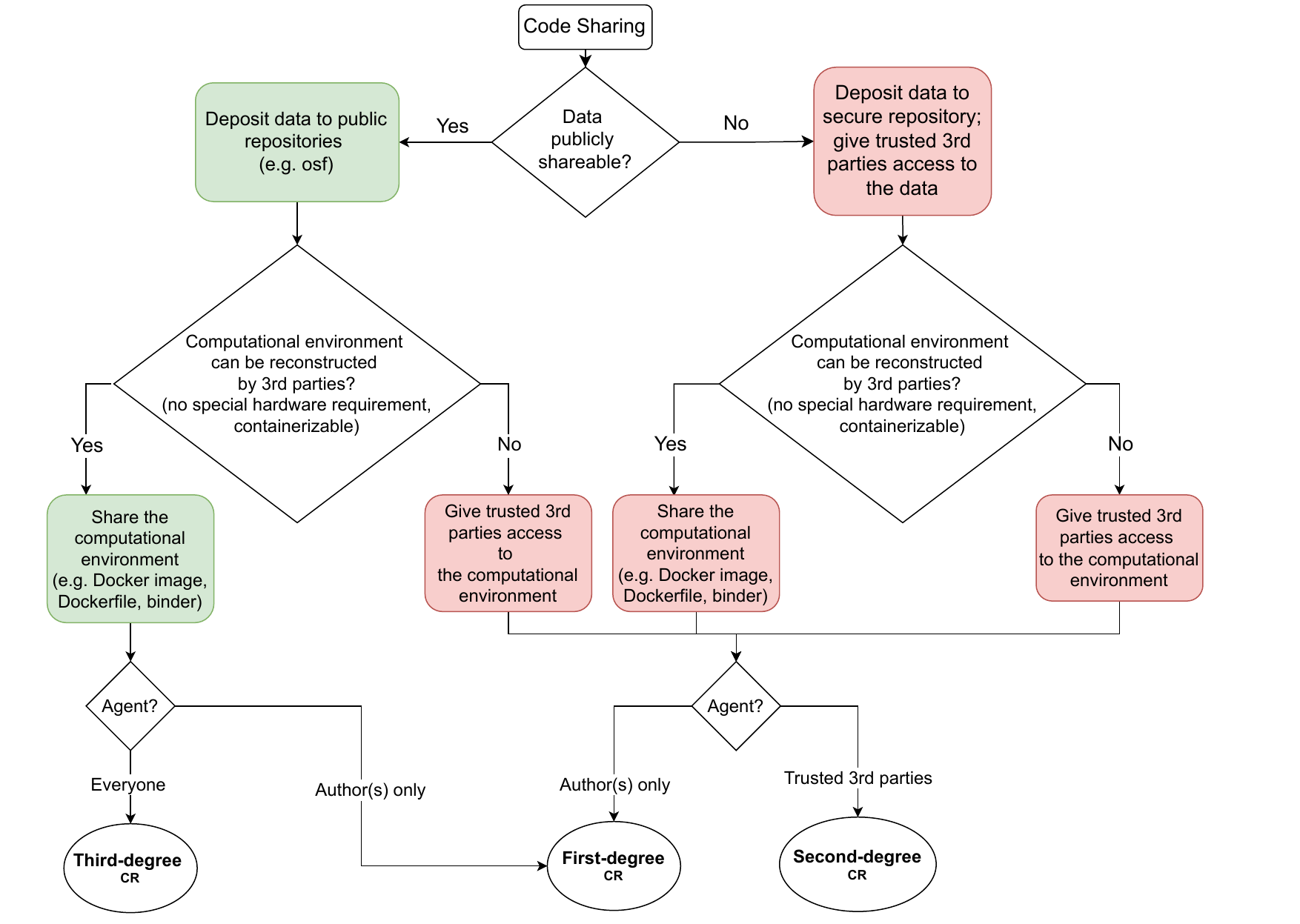}
    \label{fig:flow}
\end{figure}

For the computational environment, we restricted our discussion to the software
layer. Modern CSS research demands heavy computational resources, e.g. GPU
clusters. Inaccessible to these resources can also render CSS research
``irreproducible''. The restrictiveness dimension in our definition is still
useful (see the conditional diamond in Figure \ref{fig:flow}): The original
researcher can either allow trusted third parties to have restrictive access to
the resources to achieve 2\textdegree CR; or try to use some off-the-shelf
equipment instead to make the reproducible materials nonrestrictive to achieve
3\textdegree CR. 

To this end, we recommend the following course of action to make research
computationally reproducible. In the optimal case where both data and code can
be shared, 3\textdegree CR can be achieved by following best practice guides to
write reproducible code, using a research compendium, and by providing a
declarative description of the computational environment that rely on open
source technologies. If data cannot be shared, researchers should still follow
all mentioned practices for establishing a reproducible pipeline and either
choose a journal that allows for a reproducibility check for 2\textdegree CR or
alternatively deposit their data with a trusted third party which allows for
secure access to sensible data~\citep{arenas2019design, recker2015paving}. 

In terms of code, we argued that there is no convincing reason to not share the
code that produced the final results of a paper. Hence, if researchers are not
willing to share their code, they should be explicit about the consequences and
label their work as not computationally reproducible.

The flowchart shown in Figure \ref{fig:flow} illustrates different choices one
can make to allow different degrees of computational reproducibility. If one
would want to achieve 3\textdegree CR, the reproducible materials must be
nonrestrictive (green path). The maximum degree of computational reproducibility
one can attain with restrictive reproducible materials is 2\textdegree CR (red
path).

While implementing our suggestions at a larger scale requires change on many
institutional levels, we believe that our tier system of computational
reproducibility contributes a conceptual clarification that may improve the
scientific discourse about reproducibility. For the implementation of our
suggestions, the most obvious approach is to design a system of incentives that
fosters proactive reproducibility either through rewards or penalties. On the
one hand, making a study reproducible should not be a burden but rather an
achievement of similar quality as publishing a paper itself. The number of
papers classified as 3\textdegree CR should be an indicator that search
committees recognize and help scholars to advance their career. On the other
hand, journals could set a minimum required order of reproducibility to be
eligible for publication. Both suggestions require large structural change,
either by rethinking the ``one-dimensional credit system'' of academia or
restructuring the publication process by hiring dedicated personnel who conduct
the reproducibility checks for journals. 

\bibliographystyle{unsrtnat}

\end{document}